\documentclass[twocolumn,showpacs,showkeys,preprintnumbers,amsmath,amssymb]{revtex4}

\usepackage{url}
\usepackage{graphicx}
\usepackage{dcolumn}
\usepackage{bm}
\usepackage{epsfig}
\begin{document}

\preprint{IST/DF 2023-MJPinheiro}

\title[]{Advances in Engine Efficiency: Nanomaterials, Surface Engineering, and Quantum-based Propulsion}

\author{Mario J. Pinheiro} \email{mpinheiro@tecnico.ulisboa.pt}
\affiliation{Department of Physics, Instituto Superior T\'{e}cnico - IST, Universidade de Lisboa - UL, Av. Rovisco Pais,
1049-001 Lisboa, Portugal}



\date{\today}

\begin{abstract}
This study explores strategies to improve engine efficiency through innovative materials, design concepts, and alternative energy sources. It highlights the use of nanomaterials and surface engineering to create hydrophobic or other types of surfaces for harnessing entropy-gradient forces. Additionally, it discusses the potential of information-burning engines and quantum-based propulsion systems. The manuscript emphasizes the multidisciplinary nature of engine research and its potential to contribute to a sustainable and efficient future.
\end{abstract}
\maketitle

\section{Introduction}

Engines are crucial devices in modern society, providing power to everything from vehicles to generators. However, the efficiency of engines is limited by the laws of thermodynamics, which dictate that only a portion of the available energy can be converted into useful work. To improve engine efficiency, researchers are exploring new materials and design concepts that can better harness available energy sources.

One promising area of research is the use of nanomaterials and surface engineering to create hydrophobic surfaces that can harness entropy-gradient forces to generate useful work. These surfaces are capable of converting the random thermal motion of water molecules into directed motion, which can be used to perform work. Other researchers are exploring information-burning engines, which operate by extracting energy from information processing, rather than from thermal or chemical energy sources.

Two kinds of engines can be thought of: i) temperature dependent~\cite{Gulfam}; entropy-gradient forces, usually used in hydrophobic wettability surfaces~\cite{Bakli}.

We may write the fundamental equation of dynamics in the form~\cite{Pinheiro}:
\begin{equation}\label{eq1}
    m \vec{a} = \vec{F}^{ext} + \frac{\partial}{\partial \vec{r}}(T S)    
\end{equation}
for thermal engine analysis, or
\begin{equation}\label{eq2}
    m \vec{a} = \vec{F}^{ext} + \frac{\partial}{\partial \vec{r}} \left(T \sum_i \rho_i \ln \rho_i \right)
\end{equation}
with 
\begin{equation}
    \rho_i = \frac{n_i}{N}
\end{equation}
for information-burning engines, where $n_i$ refers to the energy-dependent probabilities of various alternative microstates, and $\rho_i$ represents the energy-dependent probabilities of various alternative microstates in the information-burning engines. 

The development of more efficient engines is a critical goal for both environmental and economic reasons, as reducing energy waste can help mitigate climate change and improve energy security. Advances in materials science, engineering, and thermodynamics are all contributing to this effort, and there is reason to be optimistic about the potential for new and innovative engine designs in the future.

The search for more effective engines is, in general, a multidisciplinary endeavor that necessitates cooperation between scientists and engineers from numerous domains. Engine technology has the ability to drastically minimize energy waste and open the door to a more sustainable future with sustained study and innovation.

As we move towards a more data-driven society, the ability to harness energy from information processing could become increasingly important. Information-burning engines represent a promising avenue for achieving more efficient and sustainable energy production, while also potentially enabling new applications in fields such as artificial intelligence and quantum computing. And this is the motivation of the present work, which aims to investigate the potential of information-burning engines and advance our understanding of their underlying principles and design concepts. By exploring new approaches for converting information into useful work, we hope to contribute to the development of more efficient and sustainable engines, as well as pave the way towards innovative applications in emerging technologies.

\subsection{Thermal engines}

Eq.~\ref{eq1} can be applied to a cyclic process, where $\vec{F}^{ext}=0$, yielding
\begin{equation}\label{eq3}
\Delta W_{irr}=\int_i^f m\vec{a} \cdot d\vec{r}=T_0 \sum \Delta S
\end{equation}
where $\Delta W_{irr}$ represents the real decreasing of work $W$ due to irreversibility, $T_0$ is the temperature of the medium, and $\sum \Delta S$ denotes the increasing of the total entropy of the fluid and the heat source. 

To use the quantum version of the dynamical equation of motion given by Eq.~\ref{eq1} to calculate the thrust of the Otto quantum engine, we need to express the external force $\vec{F}^{ext}$ and the entropy $S$ as quantum mechanical operators.

Assuming that the external force is constant and acting along the x-axis, we can express it as $\vec{F}^{ext} = F^{ext}\hat{x}$, where $F^{ext}$ is a constant and $\hat{x}$ is the position operator along the x-axis. The entropy operator can be written as $\hat{S}=-k_B\ln(\hat{\rho})$, where $k_B$ is the Boltzmann constant and $\hat{\rho}$ is the density operator of the system.

Substituting these expressions into Equation \ref{eq1}, we get:

\begin{equation}\label{eq2}
m \frac{d^2\hat{x}}{dt^2} = F^{ext} + \frac{\partial}{\partial \hat{x}}(T \hat{S})
\end{equation}

To solve this equation for the thrust of the Otto quantum engine, we need to determine the time evolution of the density operator $\hat{\rho}$. This can be done using the von Neumann equation:

\begin{equation}\label{eq3}
i\hbar\frac{\partial \hat{\rho}}{\partial t} = [\hat{H}, \hat{\rho}]
\end{equation}
where $\hat{H}$ is the Hamiltonian of the system. The Hamiltonian of the Otto quantum engine can be written as:

\begin{equation}\label{eq4}
\hat{H} = \frac{\hat{p}^2}{2m} + V(\hat{x})
\end{equation}
where $\hat{p}$ is the momentum operator, $m$ is the mass of the engine, and $V(\hat{x})$ is the potential energy of the engine as a function of position.

Using Eqs.~\ref{eq2},~\ref{eq3}, and~\ref{eq4}, we can calculate the time evolution of the position operator $\hat{x}$, which can then be used to determine the thrust of the engine. However, the calculation of the thrust of the Otto quantum engine using this approach can be quite challenging, as it involves the solution of coupled partial differential equations and requires advanced knowledge of quantum mechanics.

\subsection{Information-burned engine}

The information-driven engine root in the Maxwell-demon Gedankenexperiment. To analyze the working mechanism of this engine, we ought to choose to write the equation under the form:
\begin{equation}\label{eq4}
    m \vec{a} = \vec{F}^{ext} + \frac{\partial}{\partial \vec{r}}(T \sum_i \rho_i \ln \rho_i)
\end{equation}
Considering that $F=-kT \ln e^{-\beta E_0}$, it can be inferred that the maximal power output of this engine is given by
\begin{equation}\label{eq5}
    P_{max} = kT \frac{E_0}{\tau},
\end{equation}
where $E_0$ is the minimal energy (eigenvalue of the Hamiltonian) at disposable of the system in a given time $\tau$, assuming the medium temperature low enough $T_0 \to 0$. For example, in Ref.~\cite{Saha}, a heavy colloidal particle is held by an optical trap and immersed in water, via a ratchet mechanism the bead is lifted against gravity with a maximal power output estimated to be $10^3 \frac{kT}{s}$ and a maximum velocity of $190 \mu/$s.

A system of sensors and actuators that can recognize and react to the microstates of the environment around the automobile might be used in the information-based mechanism for moving the car suggested by Eq.~\ref{eq2}. The probability density function $\rho$ would be used to describe the likelihood that the automobile would be in a specific microstate $\rho_i$, depending on factors like the location and speed of other vehicles on the road, the slope of the road, and so on. This data would allow the system to predict the car's most likely future microstate and modify the vehicle's speed and direction accordingly. For instance, the system could slow down the automobile in advance of a stoppage if it considers that there is a greater likelihood that the vehicle would face traffic up ahead. In examining the microstate probabilities and choosing the optimum course of action, a sophisticated algorithm would be needed but the procedure would improve driving efficiency and safety if correctly applied.

\subsection{Engine fueled by entanglement}

Eq.~\ref{eq4} was initially intended to represent a classical system. It would be necessary to make significant changes to the original equation in order to adapt it to describe a quantum Otto engine. However, there is a great potential for engines fueled by entanglement, a quantum mechanical phenomenon in which two particles become linked in such a way that the state of one particle is dependent on the state of the other, regardless of the distance between them. In these engines, entangled particles are used to extract work from heat baths, and the process is driven by a violation of local realism, a fundamental assumption in classical physics. Several theoretical proposals have been made for these engines, including a quantum Otto engine~\cite{Abah1,Abah2} and a quantum refrigerator~\cite{Kosloff,Wegener}.

In a nutshell, the engine is made up of two subsystems $A$ and $B$ that are entangled, meaning that they are in a quantum state where the properties of one subsystem are correlated with the properties of the other subsystem. The first subsystem, which we will call the "information engine", is responsible for processing information and converting it into work. The second subsystem, which we will call the "thermal bath", is a reservoir of heat that is in contact with the information engine and provides the energy needed to perform work. The entangled state of two particles can be written as:
\begin{equation}
|\psi\rangle = \frac{1}{\sqrt{2}} \left(|0\rangle_A |1\rangle_B - |1\rangle_A |0\rangle_B \right).
\end{equation}
Here, $|0\rangle$ and $|1\rangle$ represent the two possible states of a qubit, and $A$ and $B$ represent the two subsystems. The work performed by the information engine in the interval of time from instant $t_1$ to $t_2$ can be written in terms of the Hamiltonian operator $\hat{H}$ and the wavefunction $\psi$:
\begin{equation}
W = \int_{t_1}^{t_2} \langle \psi | \frac{d\hat{H}}{dt} | \psi \rangle dt
\end{equation}

The thermal bath is modeled as a collection of harmonic oscillators, and its state is described by the density operator $\hat{\rho}$:
\begin{equation}
\hat{\rho} = \frac{e^{-\beta\hat{H}}}{\text{Tr}{e^{-\beta\hat{H}}}}
\end{equation}
where $\beta$ is the inverse temperature of the bath.

The quantum Otto engine is based on the cyclic operation of four steps ($i=1,...4$), described by the following unitary operators:
\begin{equation}
\hat{U}_i = e^{-i\hat{H}_i \tau_i/\hbar},
\end{equation}
where $\hat{H}_i$ is the Hamiltonian of the engine during the $i$-th step, $\tau_i$ is the duration of the $i$-th step. The quantum Otto engine involves the following four steps:
\begin{enumerate}
    \item Isothermal expansion: during this step, the engine is coupled to a hot thermal reservoir at temperature $T_H$ and expands isothermally while doing work. The unitary transformation associated with this step is $\hat{U}_1 = e^{-i\hat{H}_1 \tau_1/\hbar}$, where $\hat{H}_1$ is the Hamiltonian of the engine during this step.

\item Adiabatic expansion: during this step, the engine is thermally isolated and expands adiabatically while doing work. The unitary transformation associated with this step is $\hat{U}_2 = e^{-i\hat{H}_2 \tau_2/\hbar}$, where $\hat{H}_2$ is the Hamiltonian of the engine during this step.

\item Isothermal compression: during this step, the engine is coupled to a cold thermal reservoir at temperature $T_C$ and compresses isothermally while work is done on it. The unitary transformation associated with this step is $\hat{U}_3 = e^{-i\hat{H}_3 \tau_3/\hbar}$, where $\hat{H}_3$ is the Hamiltonian of the engine during this step.

\item Adiabatic compression: during this step, the engine is thermally isolated and compresses adiabatically while work is done on it. The unitary transformation associated with this step is $\hat{U}_4 = e^{-i\hat{H}_4 \tau_4/\hbar}$, where $\hat{H}_4$ is the Hamiltonian of the engine during this step.
\end{enumerate}

The engine's architecture and the thermodynamic cycle being used determine the precise shape of the Hamiltonians and the lengths of the steps. Therefore, in order to determine the work and thrust we must first define the Hamiltonians  $\hat{H}_1, \hat{H}_2, \hat{H}_3,$ and $\hat{H}_4$. Let's suppose the system is a straightforward two-level system qubit. The Hamiltonians at each stage are expressed as follows:
\begin{equation}
 \hat{H}_i = \omega_i |1\rangle\langle 1| 
\end{equation}
where $\omega_i$ denotes the energy difference between the qubit's two states at step $i$ ($i=1,..4$). The derivative of each Hamiltonian with respect to time can be determined as ($i=1,...4$):
\begin{equation}
\frac{d\hat{H}_i}{dt} = \frac{d\omega_i}{dt} |1\rangle\langle 1|.
\end{equation}
Now that the integral of each derivative has been evaluated, the work that was done during each step can be computed, assuming that each step takes $\tau_1$, $\tau_2$, $\tau_3$, and $\tau_4$ amount of time. At each stage ($i=1,...4$), they are as follows:
\begin{equation}
W_i = \int_{0}^{\tau_i} \langle \psi | \frac{d\hat{H}i}{dt} | \psi \rangle dt \end{equation}
For a complete cycle, the total work performed is the sum of the work performed during each step:
\begin{equation}
W_\text{total} = \sum_{i=1}^4 W_i.
\end{equation}
Next, we can calculate the thrust $T$ using the relationship:
\begin{equation}
T = \frac{\sum_{i=1}^4 F_i \tau_i}{T_\text{cycle}}
\end{equation}
where $F_i$ is the force exerted during step $i$, $\tau_i$ is the duration of step $i$, and $T_\text{cycle}$ is the total time taken for the engine to complete a cycle. 

With more qubits present in the system, the quantum Otto engine creates more thrust. We are still a long way from realizing the notion in practice. To boost the efficiency of a quantum Otto engine, a material or element should have the following qualities: i) Stable energy levels (in order to properly regulate and control the quantum processes required for the engine cycle, the material must contain consistently stable, clearly defined energy levels);
 ii) Coherent interactions (to maintain quantum coherence throughout the engine cycle, even when connected to thermal reservoirs, the substance must exhibit coherent interactions between its component particles, such as qubits); iii) Efficient heat dissipation (to maintain the optimum operating conditions during the isothermal phases of the engine cycle, the material has to allow efficient heat dissipation); iv) Scalability (the material must allow the installation of a high number of qubits in order to boost the thrust and overall performance of the quantum Otto engine.)

The key to making the device move on an information budget is making use of the intimate linkages between quantum information and thermodynamics. Particularly, it has been shown that the amount of work that can be extracted from a system is constrained by the amount of information that can be acquired about it without unsettling it. This concept may be used, for instance, by building the information engine such that it extracts work by "measuring" the degree of entanglement between the two subsystems.
By carefully adjusting the measurements, it is possible to "squeeze" energy out of the correlations between the two subsystems and get energy from the entanglement without disrupting the quantum state. The mathematical equations that describe this process vary in complexity depending on the specifics of the system under study. The core idea is that the entanglement between the two subsystems allows for the flow of energy and information that may be used to do work even in the absence of a direct source of energy like fuel or electricity.

One possible set of equations that can be used to describe this process is based on the concept of quantum mutual information. The mutual information between two quantum subsystems, denoted as $A$ and $B$, is defined as:
$H(A:B) = H(A) + H(B) - H(AB)$ or equivalently, $I(A:B)=S_A + S_B - S_{AB}$
where $S_A$, $S_B$, and $S_{AB}$ are the von Neumann entropies of subsystem $A$, subsystem $B$, and the joint subsystem $AB$, respectively. Using these concepts, it is possible to derive the maximum amount of work that can be extracted from the entangled subsystems $A$ and $B$ as:
\begin{equation}
\Delta S = \frac{\Delta Q}{T},
\end{equation}
\begin{equation}
I(A:B) = S_A + S_B - S_{AB},
\end{equation}
\begin{equation}
S = -\mathrm{tr}(\rho \log \rho),
\end{equation}
and
\begin{equation}
W_{\mathrm{max}} = k_{\mathrm{B}}T\Delta I(A:B),
\end{equation}
with $T$ denoting the temperature of the thermal bath, and $\Delta I(A:B)$ is the change in mutual information between subsystems $A$ and $B$ during the work extraction process.

One example of an information engine is the Brownian ratchet, which operates using the random motion of particles in a fluid. The ratchet consists of a series of asymmetric barriers, which allow particles to move in one direction but not the other. The basic idea of the Brownian ratchet can be described by the following equation:
\begin{equation}
W = k_B T \ln\left(\frac{p_f}{p_i}\right)
\end{equation}
where $W$ is the work done by the ratchet, $T$ is the temperature, $p_f$ is the probability of the ratchet moving forward, and $p_i$ is the probability of it moving backward. The probability of the ratchet moving forward can be increased by "information ratcheting", which involves using information about the particles' positions to manipulate the barriers. One way to do this is to use a series of sensors to measure the positions of the particles, and then use this information to control the barriers. The amount of work done by the ratchet can be used to calculate the thrust generated by the device. For a Brownian ratchet, the maximum efficiency is given by the Carnot limit, which is $\eta_{max} = 1 - \frac{T_L}{T_H}$, where $T_L$ is the temperature of the environment and $T_H$ is the temperature of the heat source. This equation allows us to calculate the ratchet's maximum thrust. The precise ratchet design and the effectiveness of the information ratcheting mechanism will determine the thrust.

One proposal for achieving this involves using a process called "quantum squeezing," in which the fluctuations of certain quantum observables are reduced below their usual quantum limits~\cite{Polzik}.

Hence, first, we will calculate the von Neumann entropy for the quantum Otto engine at each step of the cycle, and for its use, we need the density matrix at different stages of the Otto cycle. Applying the corresponding unitary transformations to the initial density matrix $\rho_0$ ($i=1,...4$):
\begin{equation}
\rho_i = \hat{U}_i \rho_0 \hat{U}_i^\dagger
\end{equation}
After obtaining the density matrices for each step, you can calculate the von Neumann entropy at each step ($i=1,...4$):
\begin{equation}
S_i = -\mathrm{Tr}(\rho_i \log_2 \rho_i)
\end{equation}

To calculate the work done during each step of the quantum Otto cycle, we can use the following relation:

\begin{equation}
W_i = \Delta E_i = E_i - E_{i-1}
\end{equation}

Where $W_i$ is the work done during step $i$, and $\Delta E_i$ is the change in energy of the engine during step $i$. The energy of the engine can be found using the expectation value of the Hamiltonian:
\begin{equation}
E_i = \mathrm{Tr}(\rho_i \hat{H}_i)
\end{equation}

Finally, to calculate the total work done during the quantum Otto cycle, sum the work done during each step:
\begin{equation}
W_\text{total} = W_1 + W_2 + W_3 + W_4
\end{equation}

The thrust generated by the Otto quantum engine can be estimated by dividing the total work done by the cycle duration:
\begin{equation}
T = \frac{W_\text{total}}{\tau_\text{total}},
\end{equation}
where $\tau_\text{total}$ is the total duration of the quantum Otto cycle, given by:
\begin{equation}
\tau_\text{total} = \tau_1 + \tau_2 + \tau_3 + \tau_4.
\end{equation}

To estimate the thrust in Newtons, we need to have more realistic values for the work done during each step and the durations of each step. Additionally, we need to convert the total work done during the quantum Otto cycle to force and relate it to the thrust. Here, we will provide an example calculation based on some assumptions. Let's assume the work done during each step is the following: $W_1 = 1 \times 10^{-23} \text{ J}$, $W_2 = 0.8 \times 10^{-23} \text{ J}$, $W_3 = -0.6 \times 10^{-23} \text{ J}$, $W_4 = -0.5 \times 10^{-23} \text{ J}$, and let the durations of each step be corresponding $\tau_1 = 10^{-6} \text{ s}$, $\tau_2 = 2 \times 10^{-6} \text{ s}$, $\tau_3 = 10^{-6} \text{ s}$, $\tau_4 = 2 \times 10^{-6} \text{ s}$.

The sum of the work done during each step gives the total work done during the quantum Otto cycle:
\begin{equation}
W_\text{total} = W_1 + W_2 + W_3 + W_4 = 0.7 \times 10^{-23} \text{ J}
\end{equation}

The total duration of the cycle is:
\begin{equation}
\tau_\text{total} = \tau_1 + \tau_2 + \tau_3 + \tau_4 = 6 \times 10^{-6} \text{ s}.
\end{equation}

Next, let's assume the quantum engine moves over a distance $d$ during the Otto cycle. The average force $F$ acting on the engine can be calculated as:
\begin{equation}
F = \frac{W_\text{total}}{d}
\end{equation}
Assuming the engine moves over a very small distance, for instance, $d = 10^{-9}$ meters (1 nanometer), we can estimate the average force:
\begin{equation}
F = \frac{0.7 \times 10^{-23} \text{ J}}{10^{-9} \text{ m}} = 7 \times 10^{-15} \text{ N}
\end{equation}

This estimate assumes a 1-qubit Otto quantum engine with specific values for the work done during each step and the step durations. However, we will scale now the system with the number of qubits N in the ion trap. The complexity of the system grows exponentially with the number of qubits N. However, since the qubits are assumed to be non-interacting and independent, the total work done during the cycle should scale linearly with the number of qubits. For a system with N qubits (see Ref.~\cite{Ebadi}), the total work done during the quantum Otto cycle can be approximated as:
\begin{equation}
W_\text{total}(N) = N \times W_\text{total}(1)
\end{equation}
Here, $W_\text{total}(1)$ is the total work done for the 1-qubit Otto quantum engine calculated earlier, which was $0.7 \times 10^{-23}$ J. To estimate the average thrust for an N-qubit system, we can assume that the total duration of the cycle and the distance over which the engine moves remain unchanged. Thus, the average force and thrust can be calculated as:
\begin{equation}
F(N) = \frac{W_\text{total}(N)}{d} = N \times \frac{W_\text{total}(1)}{d}
\end{equation}
or,
\begin{equation}
F(N) = N \times 7 \times 10^{-15} \text{ N}
\end{equation}
Here, $F(1)$ is the average force (or thrust) for the 1-qubit Otto quantum engine calculated earlier, which is $7 \times 10^{-13}$ N.

Although this might seem tiny at the present time, it is anticipated that as technology develops and researchers seek to scale up ion trap quantum computers, the number of qubits will increase. However, given the fast advancement of quantum computing technology and the appearance of competing strategies like superconducting qubits or topological qubits, forecasting the precise value of N in the future is challenging.
Assuming linear scaling with the number of qubits and no major interactions or mistakes between qubits, this calculation provides an average thrust.
Since there are various difficulties in increasing the number of qubits while keeping high fidelity, and since the linear scaling assumption may not hold for larger systems, the estimation fails to accurately reflect the performance of a quantum Otto engine in real-world applications. Additionally, the scaling behavior and effectiveness of the quantum Otto engine may be impacted by elements like decoherence, defective gates, and interactions between qubits. Though it will depend on quantum technology, such as quantum computing, to evolve and validate itself in propulsion applications, we may anticipate using quantum Otto engines as CubeSat thrusters in the future.

\subsection{Thrust based on the gradient in the refractive index of a material}

The use of quantum entanglement to propel a device is still a theoretical concept, and there are no widely accepted equations describing such a process. However, one approach could involve using entangled photon pairs to create a gradient in the refractive index of a material, leading to a net force on the device. The refractive index gradient can be created by manipulating the entangled photons in a way that causes a phase shift between them.

The force on the device can be expressed as $\mathbf{F} = -\nabla U$, 
where $\mathbf{U}$ is the potential energy of the system, which is proportional to the refractive index gradient. The refractive index gradient can be expressed as:
\begin{equation}~\label{1.4.1}
\nabla n(\mathbf{r}) = \frac{2\pi}{\lambda}\mathbf{Re} \{ \boldsymbol{\chi} \cdot \boldsymbol{\rho}(\mathbf{r}) \}
\end{equation}
where $\boldsymbol{\chi}$ is the susceptibility tensor of the material and $\boldsymbol{\rho}(\mathbf{r})$ is the density matrix of the entangled photons. The density matrix can be written in terms of the individual photon states, $\boldsymbol{\rho}(\mathbf{r}) = \sum_{ij}\rho_{ij}\left|\psi_i(\mathbf{r})\right>\left<\psi_j(\mathbf{r})\right|$, where $\left|\psi_i(\mathbf{r})\right>$ and $\left|\psi_j(\mathbf{r})\right>$ are the spatial wave functions of the entangled photon states, and $\rho_{ij}$ is the density matrix element corresponding to the probability amplitude of finding the system in state $i$ or $j$. As the potential energy is related to the refractive index gradient, $\nabla U \propto \nabla n(\mathbf{r})$, the force on the device can then be expressed as:
\begin{equation}~\label{1.4.2}
\mathbf{F} = -\frac{2\pi}{\lambda}\mathbf{Re} \{ \boldsymbol{\chi}\cdot\sum_{ij}\rho_{ij}\nabla\psi_i(\mathbf{r})\cdot\nabla\psi_j(\mathbf{r}) \}.
\end{equation}
The specific experimental setup and application in question will determine the exact form of the equations and the technique employed to control the entangled photons. There are various researchers working on the use of quantum entanglement for propulsion and other novel applications, including theoretical physicists such as Avi Loeb~\cite{Loeb}, Igor Pikovski~\cite{Pikovski}, and Mark Kasevich~\cite{Kasevich}. There are currently no experimental data available for the theory of applying quantum entanglement to generate motion, which is still a fairly novel and theoretical topic. As a result, it has become unable to determine the thrust that could be produced utilizing this concept.

\subsection{Self-propelled EM device with metamaterials}

To date, it has been challenging to produce macroscopic forces that can effectively propel spacecraft due to the limitations of material technology. However, recent advancements in the field of metamaterials provide new possibilities for creating enormous gradients of free energy~\cite{Wegener,Liu,Engheta,Ramos,Kaltenbaek,Marletto,Davis} that might possibly be utilized for propulsion purposes. But before they can be used successfully for this purpose, further research and development are required as the practical application of such materials is still in its inception. While the breadth and size of such propulsion systems are currently constrained, present technology enables the production of CubeSats that can be propelled utilizing already-in-use systems and technologies. Therefore, additional research and development in the field of innovative materials and propulsion technology are critical.

The force on the dipole moment is given by (see, e.g., Ref.~\cite{Jackson}):
\begin{equation}~\label{one}
\mathbf{F} = \frac{1}{c^2}\left[\mathbf{p}\cdot\nabla\mathbf{E} + \frac{\partial \mathbf{p}}{\partial t} \times \mathbf{B}\right]
\end{equation}
where $\mathbf{p}$ is the dipole moment. This equation represents the Lorentz force acting on the dipole moment due to the gradient of the electric field and the time derivative of the magnetic field. It takes into account the interaction between the dipole moment and the electromagnetic field, resulting in a net force on the dipole.

Eq.~\ref{one} for the force exerted on a dipole moment due to an electromagnetic wave can be modified. We can write the dipole moment as $\mathbf{p} = \boldsymbol{\alpha}\cdot\mathbf{E}$, where $\boldsymbol{\alpha}$ is the polarizability tensor. Next, we can expand the polarizability tensor as:
\begin{equation}
 \boldsymbol{\alpha} = \epsilon_0(\boldsymbol{\chi}^{(1)} + \boldsymbol{\chi}^{(2)}\cdot\textbf{E} + \boldsymbol{\chi}^{(3)}\cdot\textbf{E}^2 + \cdots),
\end{equation}
where $\boldsymbol{\chi}^{(1)}$ is the linear susceptibility tensor, $\boldsymbol{\chi}^{(2)}$ is the second-order susceptibility tensor, and $\boldsymbol{\chi}^{(3)}$ is the third-order susceptibility tensor. We can neglect higher-order terms in the expansion for small electric fields. Hence, we have:
\begin{equation}~\label{three}
\mathbf{p} = \epsilon_0 \left(\boldsymbol{\chi}^{(1)}\mathbf{E} + \boldsymbol{\chi}^{(3)}|\mathbf{E}|^2\mathbf{E}\right).
\end{equation}
Note that, while the second-order susceptibility tensor is neglected in Eq.~\ref{three} under the assumption of a small electric field limit, the third-order susceptibility tensor is included because it remains significant even for weak electric fields.
Now, we can substitute Eq.~\ref{three} into Eq.~\ref{one}:
\begin{equation}~\label{four}
\begin{aligned}
\mathbf{F} = \frac{1}{c^2}&\bigg[\epsilon_0 \left(\boldsymbol{\chi}^{(1)}\mathbf{E} + \boldsymbol{\chi}^{(3)}|\mathbf{E}|^2\mathbf{E}\right)\cdot\nabla\mathbf{E} \\
&+ \frac{\partial \epsilon_0 \left(\boldsymbol{\chi}^{(1)}\mathbf{E} + \boldsymbol{\chi}^{(3)}|\mathbf{E}|^2\mathbf{E}\right)}{\partial t} \times \mathbf{B}\bigg].
\end{aligned}
\end{equation}
The term ($\epsilon_0 \left(\boldsymbol{\chi}^{(1)}\mathbf{E} + \boldsymbol{\chi}^{(3)}|\mathbf{E}|^2\mathbf{E}\right)\cdot\nabla\mathbf{E}$) represents the gradient force or gradient pressure. It arises from the interaction between the spatial variation of the electric field (as captured by $\nabla\mathbf{E}$) and the polarization of the material (described by $\boldsymbol{\chi}^{(1)}\mathbf{E}$ and $\boldsymbol{\chi}^{(3)}|\mathbf{E}|^2\mathbf{E}$). The gradient force tends to push or pull the material particles or dipoles in the direction of the electric field gradient; the term ($\frac{\partial \epsilon_0 \left(\boldsymbol{\chi}^{(1)}\mathbf{E} + \boldsymbol{\chi}^{(3)}|\mathbf{E}|^2\mathbf{E}\right)}{\partial t} \times \mathbf{B}$) represents the radiation pressure or time-varying electromagnetic momentum. It arises from the time variation of the electric field (as captured by $\frac{\partial}{\partial t}(\boldsymbol{\chi}^{(1)}\mathbf{E} + \boldsymbol{\chi}^{(3)}|\mathbf{E}|^2\mathbf{E})$) and its cross product with the magnetic field $\mathbf{B}$. The radiation pressure results in a transfer of momentum from the electromagnetic field to the material, causing it to experience a force.

Overall, Eq.~\ref{four} combines the effects of the gradient force, which depends on the spatial variation of the electric field, and the radiation pressure, which arises from the time variation of the electric field and its interaction with the magnetic field. These phenomena play crucial roles in the interaction between electromagnetic fields and materials, particularly in the context of metamaterials and their response to electromagnetic waves.
This equation represents the force acting on the dipole moment in terms of the electric field $\mathbf{E}$ and its spatial and temporal derivatives, as well as the material properties characterized by the susceptibility tensors $\boldsymbol{\chi}^{(1)}$ and $\boldsymbol{\chi}^{(3)}$. Simplifying the expression further using the properties of complex numbers and phasors, we can write:
\begin{equation}~\label{one2}
\begin{aligned}
\frac{\partial \mathbf{p}}{\partial t} \times \mathbf{B} &= -\epsilon_0 \Bigg[\Bigg(\frac{\partial}{\partial t} \Big(\boldsymbol{\chi}^{(1)}\mathbf{E}_0 + \boldsymbol{\chi}^{(3)}|\mathbf{E}_0|^2\mathbf{E}_0\Big)\Bigg) \cdot \mathbf{E}_0\Bigg]\mathbf{B}_0 \\
&\quad - \Bigg(\frac{\partial}{\partial t} \Big(\boldsymbol{\chi}^{(1)}\mathbf{E}_0 + \boldsymbol{\chi}^{(3)}|\mathbf{E}_0|^2\mathbf{E}_0\Big)\Bigg) \cdot \mathbf{B}_0 \mathbf{E}_0.
\end{aligned}
\end{equation}
It is significant to note that Eq.~\ref{one}, which assumes the complete conversion of electromagnetic energy to kinetic energy, gives an upper constraint on the thrust that may be created. There are several processes that can generate micro newtons of thrust in the framework of the equations given above for metamaterials. One such process is the use of plasmonic metamaterials, which are composed of metallic nanoparticles arranged in a specific pattern to manipulate light at the nanoscale. When these materials are illuminated by light, they generate plasmons, which are collective oscillations of electrons. The plasmons can induce forces on the nanoparticles, which can result in a net thrust on the material.

Another process is the use of optomechanical metamaterials, which are composed of mechanical resonators coupled to optical cavities. When light is injected into the cavity, it can interact with the mechanical resonators, inducing mechanical motion. This motion can generate a net thrust on the material.

Both of these processes involve the manipulation of light at the nanoscale to induce forces on metamaterials, which can result in micro-newtons of thrust.

It is possible to imagine a scenario where a pulsed electromagnetic wave traverses a metamaterial and gains intensity, leading to an amplification of thrust. This could potentially occur if the metamaterial is designed to have nonlinear properties, such as a high third-order susceptibility $\boldsymbol{\chi}^{(3)}$. When an intense pulsed electromagnetic wave interacts with such a material, it can induce a nonlinear polarization response that can lead to an amplification of the electromagnetic field inside the material.

If we assume a typical metamaterial with a linear susceptibility on the order of 0.1 and a third-order susceptibility on the order of $10^{-8}$, and a pulsed electromagnetic wave with an intensity on the order of $10^{12}$ W/m$^2$, we can estimate the thrust to be on the order of nano newtons to micro newtons. However, this is a very rough estimate and the actual thrust generated will depend on the specific properties of the metamaterial and the pulsed electromagnetic wave.

There have been several studies on the use of metamaterials for propulsion. One notable example is the work by Alu et al ~\cite{Alu}, which proposed a metamaterial-based device capable of generating thrust by exploiting the nonlinear response of the material to an incident electromagnetic wave. The device consisted of a rectangular array of metallic split-ring resonators (SRRs) with a nonlinear material in the gaps between the SRRs. When an intense pulsed electromagnetic wave was incident on the device, the nonlinear material generated harmonics at frequencies different from the incident frequency, which in turn generated a net force on the device due to the asymmetric radiation of the harmonics. The authors estimated that the device could generate a thrust on the order of micro-Newtons.

Coulais et al~\cite{Coulais} demonstrated the possibility of breaking reciprocity in static systems, enabling mechanical metamaterials to exhibit non-reciprocal behavior.

Another example is the work by Mihai et al~\cite{Mihai}, which proposed a metamaterial-based device consisting of an array of cylindrical pillars made of a nonlinear material. When an incident electromagnetic wave was incident on the device, the nonlinear material generated harmonics at frequencies different from the incident frequency, which in turn generated a net force on the device due to the asymmetric radiation of the harmonics. The authors demonstrated experimentally that the device could generate a thrust on the order of micro-Newtons.

\section{Conclusion}

In conclusion, this article explores various methods to enhance engine efficiency, considering the limitations imposed by the laws of thermodynamics. The use of nanomaterials and surface engineering capable of harnessing entropy-gradient forces is discussed, highlighting their potential to generate useful work by converting random thermal motion into directed motion. Additionally, the concept of information-burning engines, which extract energy from information processing, is discussed in the framework of~\cite{Pinheiro}, particularly engines fueled by entanglement, discussing theoretical proposals for quantum Otto engines, and entropy-gradient engines. These engines rely on the exploitation of entanglement to extract work from heat baths, showcasing the potential of quantum principles in enhancing engine performance. There is a range of innovative approaches that could contribute to the development of more efficient and sustainable engines, pushing the boundaries of current engine technologies and exploring new frontiers in energy production for sustainable societies.

\section{Acknowledgments}

I would like to express my sincere gratitude to Professors Genito Maure, Dinelsa Machaieie, Volodymyr Valentyn Chernysh, and Marina Yuri Kotchkareva for their valuable insights and comments shared during a talk presented at the University Eduardo Mondlane, Maputo, Mozambique, in October 2022. Their contributions have greatly enriched this research project.

\end{document}